\documentclass[a4paper,fleqn,usenatbib]{mnras}

\usepackage{newtxtext,newtxmath}

\usepackage[T1]{fontenc}
\usepackage{ae,aecompl}

\usepackage{graphicx}	
\usepackage{amsmath}	
\usepackage{amssymb}	
\usepackage{flushend}

\title[The Nature of the 2016 $\gamma$-ray Emission from 1749+096]{Exploring the Nature of the 2016 $\gamma$-ray Emission in the Blazar 1749+096}

\author[D.-W. Kim et al.]{
Dae-Won Kim$^{1}$,
Sascha Trippe$^{1}$\thanks{Corresponding author: trippe@astro.snu.ac.kr},
Sang-Sung Lee$^{2,3}$,
Jae-Young Kim$^{4}$,
\newauthor{Juan-Carlos Algaba$^{1}$,}
Jeffrey Hodgson$^{2}$,
Jongho Park$^{1}$,
Motoki Kino$^{5,6}$,
\newauthor{Guang-Yao Zhao$^{2}$,}
Kiyoaki Wajima$^{2}$,
Jee Won Lee$^{2}$
and Sincheol Kang$^{2,3}$
\\
$^{1}$Department of Physics and Astronomy, Seoul National University, Gwanak-gu, Seoul 08826, Korea; dwkim@astro.snu.ac.kr\\
$^{2}$Korea Astronomy and Space Science Institute, 776 Daedeok-daero, Yuseong-gu, Daejeon 34055, Korea\\
$^{3}$Korea University of Science and Technology, 217 Gajeong-ro, Yuseong-gu, Daejeon 34113, Korea\\
$^{4}$Max-Planck-Institut f\"{u}r Radioastronomie, Auf dem H\"{u}gel 69, 53121 Bonn, Germany\\
$^{5}$National Astronomical Observatory of Japan, 2211 Osawa, Mitaka, Tokyo 1818588, Japan\\
$^{6}$Kogakuin University, Academic Support Center, 2665-1 Nakano, Hachioji, Tokyo 192-0015, Japan\\
}

\date{Accepted XXX. Received YYY; in original form ZZZ}

\pubyear{2018}

\begin{document}
\label{firstpage}
\pagerange{\pageref{firstpage}--\pageref{lastpage}}
\maketitle

\begin{abstract}
Recent \emph{Fermi}-Large Area Telescope (LAT) light curves indicate an active $\gamma$-ray state spanning about five months from 2016 June to 2016 October in the BL Lac object 1749+096 (OT\,081). During this period, we find two notable $\gamma$-ray events: an exceptionally strong outburst followed by a significant enhancement (local peak). In this study, we analyze multi-waveband light curves (radio, optical, X-ray, and $\gamma$-ray) plus very-long baseline interferometry (VLBI) data to investigate the nature of the $\gamma$-ray events. The $\gamma$-ray outburst coincides with flux maxima at longer wavelengths. We find a spectral hardening of the $\gamma$-ray photon index during the $\gamma$-ray outburst. The photon index shows a transition from a softer-when-brighter to a harder-when-brighter trend at around 1.8 $\times$ $10^{-7}$ ph cm$^{-2}$ s$^{-1}$. We see indication that both the $\gamma$-ray outburst and the subsequent enhancement precede the propagation of a polarized knot in a region near the VLBI core. The highest polarized intensity, 230\,mJy, and an electric vector position angle rotation, by $\sim$32$^{\circ}$, are detected about 12 days after the $\gamma$-ray outburst. We conclude that both $\gamma$-ray events are caused by the propagation of a disturbance in the mm-wave core.
\end{abstract}



\section{Introduction}

Since the dawn of the \emph{Fermi}-LAT era it has became clear that blazars are the dominant source type ($\sim$75\% of sources, excluding unknown blazar types) in the extragalactic $\gamma$-ray sky above 10 GeV \citep{Ajello2017}. The spectral energy distributions (SED) of blazars are supposedly dominated by the emission from relativistic jets \citep{Lewis2018}. In leptonic models, blazar jets radiate from radio to $\gamma$-ray via two primary mechanisms, synchrotron and inverse Compton scattering (IC); these are responsible for radio to UV and X-ray to $\gamma$-ray emission, respectively (but see also \citealt{Cerruti2015}, for hadronic models explaining the high energy emission via proton synchrotron in the jets). \citet{Fossati1998} found a correlation (the `blazar sequence') in blazar SEDs between the synchrotron hump and the IC hump (e.g., \citealt{KimDW2017}) which implies a tight connection between the powerful radio jet and $\gamma$-ray emission in blazars \citep{Hughes2011, Jorstad2013}. Indeed, several studies suggest a strong correlation between the radio and $\gamma$-ray emission \citep{Tavares2012, Lico2017}. However, the origin and physical radiation mechanisms of the high energy ($\gamma$-rays) emission in blazar jets are still a matter of debate \citep{Chatterjee2012, Fuhrmann2014, Moerbeck2014, Jorstad2016, Lewis2018}.

Emission regions located within relativistic jets (e.g., \citealt{Marscher2008, Boccardi2017}) are thought to be the production site of $\gamma$-radiation -- notably, of $\gamma$-ray flares -- in blazars. Models of the location of $\gamma$-ray flares broadly suggest two potential locations (e.g., \citealt{Dotson2012}): the broad line region (BLR) and the radio core which is the surface of unity optical depth. At sub-parsec scale distances from the black hole (BH), optical--UV photons from the BLR can be up-scattered by electrons accelerated in a relativistic shock \citep{Sikora1994}. Such shocks are supposed to occur when a disturbance moving along the jet passes through the acceleration and collimation zone (ACZ) (see also \citealt{Marscher2008}, for the canonical model of a blazar jet). In opposition to this scenario, many observations which found concurrence of events at different wavebands (including $\gamma$-rays) point to the radio core (or a region downstream of the core) as the place of origin of $\gamma$-ray events. In this scenario, $\gamma$-radiation is produced via inverse Compton scattering of radio-to-IR seed photons from the jet itself (e.g., \citealt{Marscher2010}) or from the dusty torus located a few parsecs from the black hole \citep{Tavares2011}. The parsec-scale scenario has been supported by \citet{Jorstad2001a, Jorstad2001b} who observed a connection of the $\gamma$-ray emission with the ejection of (apparently) superluminal jet components and the flux density of the VLBI core. Further insights are provided multi-waveband photometry plus polarimetry VLBI campaigns. \citet{Agudo2011} reported that a disturbance propagating through the 7-mm core caused a very strong $\gamma$-ray outburst with counterparts at frequencies from radio to $\gamma$-rays in the blazar AO 0235+164 in 2008. The disturbance showed strong linear polarization, corresponding to the signature of a moving shock \citep{Marscher2010, Wehrle2012, Jorstad2013}. \citet{Lico2014} found a correlation between the flux of the radio core (at 15, 24, and 43 GHz) and $\gamma$-ray emission in the blazar Mrk\,421 throughout 2011, albeit with substantial scatter (with Pearson correlation coefficients of 0.42 to 0.46). During their observations, only the first $\gamma$-ray peak (on MJD 55627) had a radio counterpart (on MJD 55621) in the 7-mm core flux. Given the sampling interval of the radio observations, it is unclear where the radio counterpart peaked. They also reported no spectral hardening in the $\gamma$-ray spectrum during the enhanced $\gamma$-ray state in Mrk\,421, which is a rare feature in blazars.

The BL Lac object 1749+096 (OT\,081, redshift 0.32, image scale 4.64 pc/mas, assuming $H_{0}$ = 71 km~Mpc~s$^{-1}$, $\Omega_{\Lambda}$ = 0.73, and $\Omega_{m}$ = 0.27) is a flat spectrum radio source emitting variable radio radiation in total intensity and linear polarization \citep{Stickel1988, Gabuzda1996}. 1749+096 has been classified as a low-synchrotron peaked (LSP) blazar,\footnote{\url{http://www.physics.purdue.edu/MOJAVE/sourcepages/1749+096.shtml}}
has been observed at X-rays, but was not detected at $\gamma$-rays before the advent of \emph{Fermi} \citep{Sambruna1999}. A review of the physical characteristics of this highly compact radio source can be found in \citet{Lu2012}, covering features such as multi-frequency variability from radio to X-ray, a quiescent flux level of below 1 Jy at high radio frequencies (above 37\,GHz), a curved extended jet, and superluminal motion of jet components, with apparent speeds from $5c$ to $21c$. \citet{Jorstad2017} presented a recent estimate of Doppler factor of $\sim$17.7 and viewing angle of $\sim$2.4$^{\circ}$ in the jet of 1749+096. The first $\gamma$-ray detection of 1749+096 was reported by \citet{Abdo2009}. Interestingly, there were no $\gamma$-ray flares until 2015, which is why the $\gamma$-ray outburst in 2016 is notable. \citet{osullivan2009} revealed the linear polarization of the 1749+096 jet at 4.6\,--\,43\,GHz by using very long baseline array (VLBA) observations. They found that the radio core shows a degree of linear polarization of about 3\% across the range of their frequencies, with the polarization angle being about $-50^{\circ}$ at 43\,GHz. Additionally, 1749+096 is known to show a Faraday rotation measure (RM) significantly different from zero at cm-wave bands \citep{Pushkarev2001}. Contrary to this, however, \citet{osullivan2009} found no significantly non-zero RM at frequencies up to 43\,GHz in the radio core during a flare, thus indicating that the underlying magnetic field is most likely responsible for EVPA changes in some specific circumstances \citep{Homan2002}. Recently, it was also found that 1749+096 shows variability in optical polarization on time scales of a few days \citep{Uemura2017}.

In this study, we explore the powerful $\gamma$-ray outburst in 1749+096 that occurred in the middle of 2016 by using multi-waveband observations including, especially, VLBI data. Overall, the multi-waveband data span about two years (2015 and 2016) across a frequency range from radio to $\gamma$-rays obtained from the Korean VLBI Network (KVN) at 22, 43, 86, and 129\,GHz; the Owens Valley Radio Observatory (OVRO) at 15\,GHz; the VLBA at 43\,GHz; the All-Sky Automated Survey for Supernovae (ASAS-SN) in the optical V-band; \emph{Swift}-XRT at X-rays; and \emph{fermi}-LAT at $\gamma$-rays. Due to a rather spotty $\gamma$-ray light curve, we focus on a specified $\gamma$-ray active period spanning $\sim$5 months (see Figure~\ref{fig:f1}) which includes both the $\gamma$-ray outburst and a notable local peak (temporary flux enhancement). We address multi-waveband correlations, the evolution of the $\gamma$-ray spectrum, and the linear polarization at 43\,GHz as observed by the VLBA. We discuss the connection between the $\gamma$-ray events and radio core activity, assuming that the primary candidate of the $\gamma$-ray production site is the radio core.

\section{Observations and Data}

\subsection{KVN 22/43/86/129 GHz \& VLBA 43 GHz}
\label{sec:obsA} 

We obtained multi-frequency VLBI data from the Interferometric Monitoring of Gamma-ray Bright AGNs (iMOGABA) project.\footnote{\url{http://radio.kasi.re.kr/sslee}} iMOGABA employs the KVN for multi-frequency simultaneous observations at 22, 43, 86, and 129\,GHz in single polarization (LCP). The KVN consists of three identical (diameter of 21\,m) antennas with baseline lengths up to $\sim$470\,km; accordingly, angular resolutions are on the order of a few milliseconds of arc. iMOGABA has been monitoring $\sim$30 $\gamma$-ray bright AGNs monthly since late 2012 \citep[see, e.g.,][]{Lee2013, Algaba2015}. Data reduction was conducted with the KVN pipeline \citep{Hodgson2016} which applies all standard procedures required for reduction of VLBI data. We used the frequency phase transfer (FPT) technique \citep{Zhao2018} to improve the quality of the data from the higher frequency bands. We followed the procedure used in \citet{Lee2016} for imaging our data with the software package \texttt{Difmap} \citep{Shepherd1997}. We conservatively estimated an error of 10\% on the flux density of each image component; for our 129\,GHz data, we applied a 30\% error due to possible systematic amplitude losses \citep{KimDW2017}. Usually, we detected only one component (i.e., the KVN core) at the map center over the four frequencies owing to the relatively large beam size of the KVN and its limited sensitivity. In a few cases, closure phase analysis at 43 and 86\,GHz made it possible to detect a jet pointing toward the northeast -- which is consistent with the known morphology of the radio jet of 1749+096 (e.g., \citealt{Lu2012}). Given the performance and limitations of the KVN, we consider those detections marginal and use only the KVN core in this study.

We selected seven VLBA observations (2016 June to 2016 November) around the time of the two $\gamma$-ray events (see Figure~\ref{fig:f1}) from the Boston University blazar group (BU) archival dataset\footnote{\url{http://www.bu.edu/blazars/VLBA_GLAST/1749.html}} to look into the source more deeply, including the linear polarization at 43\,GHz. The BU group has been monitoring several tens of $\gamma$-ray bright blazars monthly via the VLBA in close association with the \emph{Fermi}-LAT \citep{Jorstad2016}. The public BU data were already fully calibrated as described in \citet{Jorstad2017}. Hence, we simply used the calibrated visibility data to produce Stokes $I$, $Q$, and $U$ maps. We imaged the data with \texttt{Difmap} and produced linear polarization maps using the Astronomical Image Processing System (AIPS) task \texttt{COMB} \citep{Moorsel1996}. We note that the BU observations missed the SC and KP stations on 2016 September 5 and the HN and MK stations on 2016 October 6. Hence, results from those epochs need to be interpreted with care. We fit circular Gaussian profiles to the image components in the total intensity maps to investigate the evolution of their flux densities, assuming again a conservative error of 10\%.

\subsection{OVRO 15 GHz}
\label{sec:obsB} 

We collected 15\,GHz data of 1749+096 from the OVRO 40\,m telescope monitoring program \citep{Richards2011}. In close association with the \emph{Fermi}-LAT program, the OVRO has been monitoring more than 1800 blazars about twice per week since 2008. The large sample size and the high cadence allow for a detailed exploration of blazar variability at cm-wavelengths. Details of the data reduction process can be found in \citet{Richards2011}. The calibrated OVRO data is available via the OVRO Internet database.\footnote{\url{http://www.astro.caltech.edu/ovroblazars/index.php?page=home}} In this study, we use OVRO flux data spanning from the beginning of 2015 to early 2017.

\subsection{ASAS-SN}
\label{sec:obsC} 

We obtained optical V-band data from the All-Sky Automated Survey for Supernovae (ASAS-SN) project\footnote{\url{http://www.astronomy.ohio-state.edu/asassn/index.shtml}} \citep{Shappee2014, Kochanek2017}. The survey is ongoing every night with 20 telescopes located around the globe including Hawaii, Chile, and South Africa. The ASAS-SN aims to survey and discover bright transients down to a V-band magnitude of about 17 across the entire sky. The project provides an online tool that produces an aperture photometry light curve for an arbitrary point on the celestial sphere, thus making it possible to study sources other than supernovae. We extract optical light curve of 1749+096 by using this online tool.

\subsection{Swift-XRT}
\label{sec:obsD} 

We collected X-ray data (0.3--10 keV) from \emph{Swift}-XRT observations \citep{Gehrels2004}. The \emph{Swift}-XRT is a mission launched in 2004 to investigate X-ray afterglows of $\gamma$-ray bursts (GRB). The UK Swift Science Data Centre\footnote{\url{http://www.swift.ac.uk/index.php}} provides an automatic online pipeline that produces high level XRT products for non-GRBs with the software package HEASOFT v6.22. We employ the online pipeline to generate an X-ray light curve of 1749+096 with a $3\sigma$ cutoff. Details of the pipeline and the data reduction process are provided by \citet{Evans2007}.

\begin{figure*}
 \includegraphics[width=15cm]{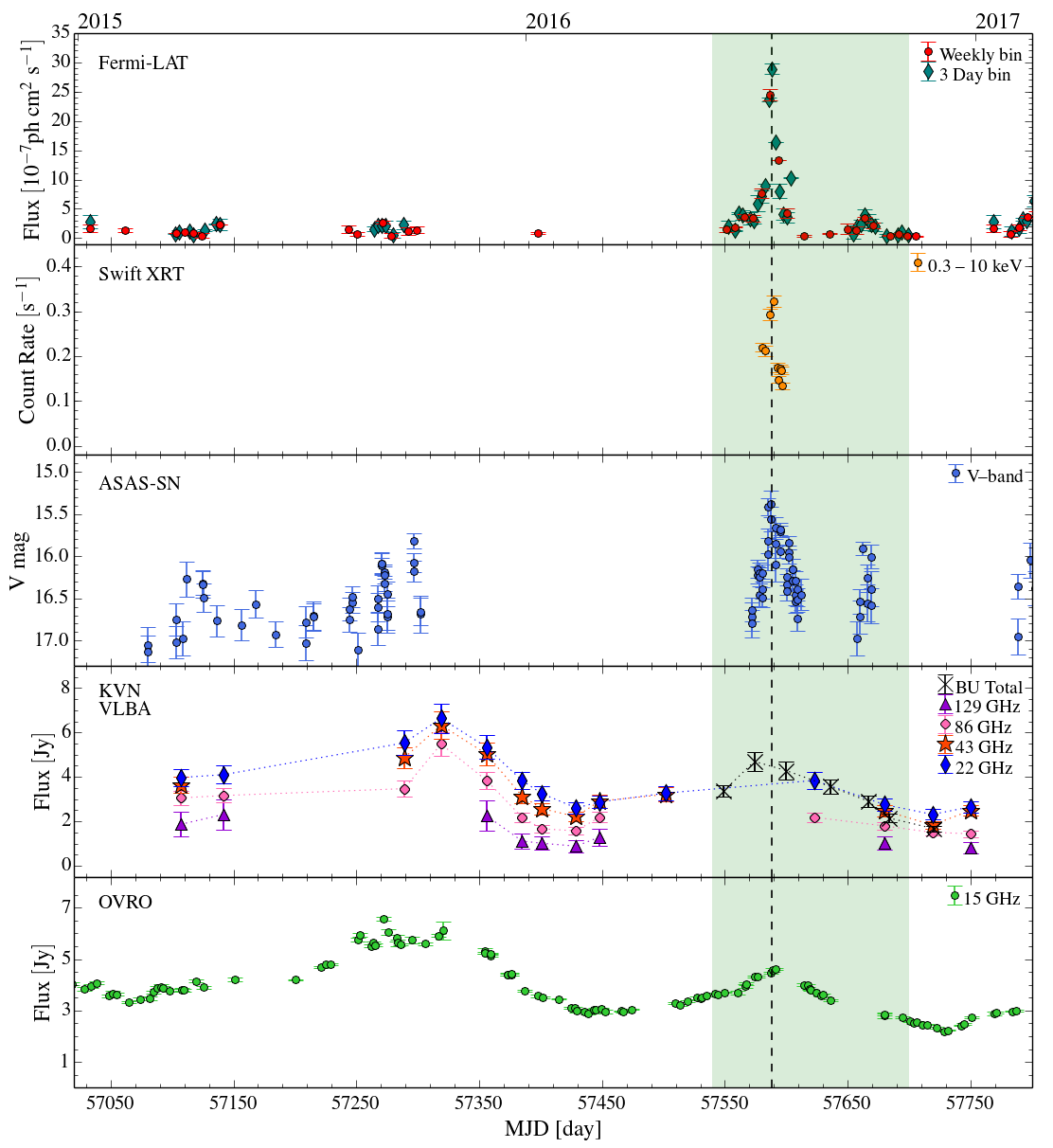}
 \caption{Multi-waveband light curves of 1749+096. From top to bottom: \emph{Fermi}-LAT at 0.1--300\,GeV, \emph{Swift}-XRT at 0.3--10\,keV, ASAS-SN at optical V-band, KVN (iMOGABA) at 22/43/86/129\,GHz plus  VLBA (BU) at 43\,GHz, and OVRO at 15\,GHz. The data spans the period from 2014 December 29 to 2017 February 16. The light green shaded region indicates the $\gamma$-ray active period (MJD 57540--57700). The black dashed vertical line indicates the 2016 July 19 (MJD 57588) $\gamma$-ray outburst.}
 \label{fig:f1}
\end{figure*}

\subsection{Fermi-LAT}
\label{sec:obsE} 

The \emph{Fermi}-LAT $\gamma$-ray space mission was launched in 2008 June to explore the high energy sky \citep{Atwood2009}. The LAT is designed to cover the energy range of 20\,MeV--300\,GeV, and performs an all-sky survey with its large field of view (2.4 sr). We use the \emph{Fermi} software \texttt{ScienceTools v10r0p5} and the instrument response function (IRF) \texttt{P8R2\_SOURCE\_V6} to extract light curves of 1749+096 from the raw LAT data. We essentially follow the data reduction steps employed by \citet{Prince2017}. The initial search radius was set to $20^{\circ}$ around 1749+096. We selected events in the SOURCE class (Pass 8) with an energy range of 0.1--300 GeV. To exclude atmospheric background events (i.e., contamination from the Earth limb $\gamma$-radiation) and select good time invervals (GTIs), we applied the \texttt{zmax} option in \texttt{gtltcube} (\texttt{zmax=90$^{\circ}$}) plus the filter \texttt{DATA\_QUAL==1 \&\& LAT\_CONFIG==1} which is the currently recommended procedure. We extracted source models within the search window from the third \textit{Fermi}-LAT catalog (3FGL). We set the spectral parameters for sources within and outside the Region of Interest (ROI) of 10$^{\circ}$ to be free and fixed to the catalog values, respectively. A power-law (PL) function was applied to the photon spectra of 1749+096. We performed an unbinned likelihood analysis where the significance of the $\gamma$-ray flux is evaluated by maximum likelihood (ML) test statistics (e.g., \citealt{Paliya2015}). We modelled the contribution by diffuse background sources with the recent isotropic background model \texttt{iso\_P8R2\_SOURCE\_V6\_v06} and the galactic diffuse emission model \texttt{gll\_iem\_v06}. We use the Perl script \texttt{like$\_$lc.pl}\footnote{\url{https://fermi.gsfc.nasa.gov/ssc/data/analysis/user/}} written by R. Corbet to produce $\gamma$-ray light curves. A weekly $\gamma$-ray light curve of 1749+096 is generated using the criterion \texttt{TS=9} (corresponding to a $3\sigma$ cutoff), flux values below this threshold are rejected. We also provide 3-day binned $\gamma$-ray light curve for further analysis. During the photon index analysis, we noted and rejected a few outliers (three and two data points in the weekly and 3-day binned data, respectively) deviating by more than $2\sigma$ from the weekly photon index trend. All relevant files and data are provided by the \emph{Fermi} data web site.\footnote{\url{https://fermi.gsfc.nasa.gov/ssc/data/access/}}

\section{Results and Analysis}

\subsection{Multi-waveband light curves}
\label{sec:resA} 

Figure~\ref{fig:f1} shows the multi-waveband light curves spanning from 2014 December 29 to 2017 February 16 (MJD 57020--57800). Until mid-2016, 1749+096 is $\gamma$-ray quiet with fluxes $\lesssim2\times10^{-7}$ ph cm$^{-2}$ s$^{-1}$ and detected only occasionally, whereas radio observations find enhanced activity peaking around mid-2015. In 2016 July, a powerful $\gamma$-ray outburst occurs, rising to about $\sim$15 times the quiescent level within 36 days (in the 3-day binned data). The $\gamma$-ray flux peaked at $2.9\times10^{-6}$ ph cm$^{-2}$ s$^{-1}$ on 2016 July 19 (MJD\,57588$\pm$1.5\,d; 3-day binned data). Counterparts to this $\gamma$-ray event can be found at all wavebands (radio, optical, and X-ray). The X-ray, optical, and cm-wave (OVRO 15\,GHz) counterparts peaked on 2016 July 20 (MJD\,57589), 2016 July 18 (MJD\,57587), and 2016 July 22 (MJD\,57591) respectively, meaning that the X-ray and optical counterparts were simultaneous with the $\gamma$-ray outburst within the error of $\pm$1.5\,day given by the time resolution of the $\gamma$-ray light curve. The 15-GHz peak occurs $\sim$3 days after the $\gamma$-ray peak; the difference might actually be larger because the OVRO light curce shows a gap right after its apparent maximum on MJD 57591. In addition to the 2016 July the $\gamma$-ray outburst, a smaller temporary $\gamma$-ray flux enhancement occurred on 2016 October 2 (MJD 57663$\pm$1.5 d), reaching up to about $3.9\times10^{-7}$ ph cm$^{-2}$ s$^{-1}$. This event appears to have no counterparts at radio wavelengths.

Unfortunately, the available information at mm-wavelengths is poor during the period we specify as `$\gamma$-ray active' from 2016 June 1 to 2016 November 8 (the period indicated in Figure~\ref{fig:f1}). However, we do see a radio counterpart to the $\gamma$-ray outburst at mm-wavelengths from the BU data; due to the rather large sampling intervals of the BU data, $\sim$1\,month, it is unclear where the mm-wave light curve peaks exactly. From the iMOGABA and OVRO light curves, we find a period of enhanced mm-radio flux in mid-2015. Contrary to the subsequent radio flare in the middle of 2016, we do not find a corresponding increase in $\gamma$-ray activity. Overall, 1749+096 shows rather quiescent, and frequently undetectable, $\gamma$-ray emission during most of our observations except the $\gamma$-ray active period in 2016. Thus, we focus on this period in the further analysis.

\begin{figure}
 \includegraphics[width=\columnwidth]{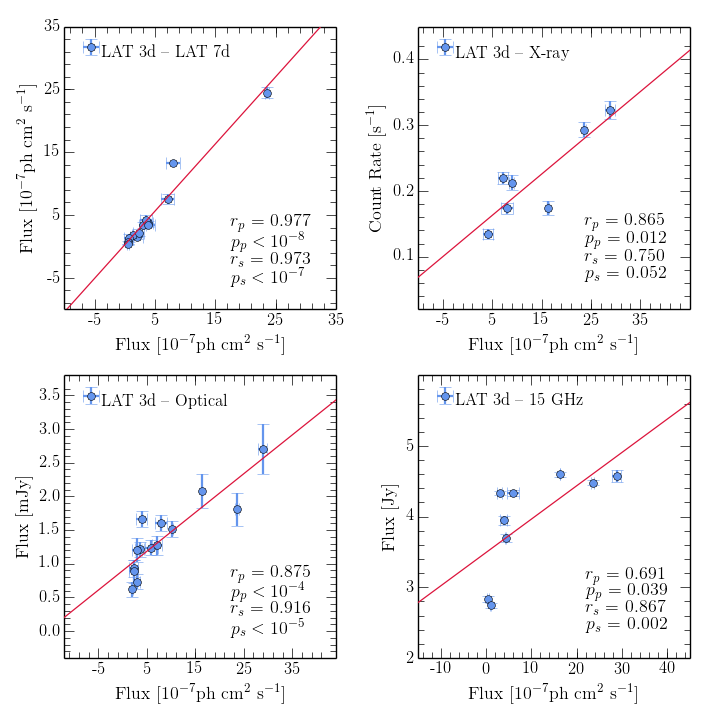}
 \caption{Correlations between $\gamma$-ray flux (3-day binning) and three other wavebands (X-ray, optical, and 15-GHz radio) for the time of high $\gamma$-ray activity in mid-to-end 2016. The top left panel shows the correlation between 3-day and 7-day $\gamma$-ray light curves as a consistency check. Correlations are tested with data points that are simultaneous within the bin size (i.e., $\pm1.5$ days) of the $\gamma$-ray data. Each panel gives Pearson ($r_p$) and Spearman rank ($r_s$) correlation coefficients values together with the corresponding false alarm probabilities ($p$ values). Red lines indicate the best-fit linear relationships.}
 \label{fig:f2}
\end{figure}

\begin{figure}
 \includegraphics[width=\columnwidth]{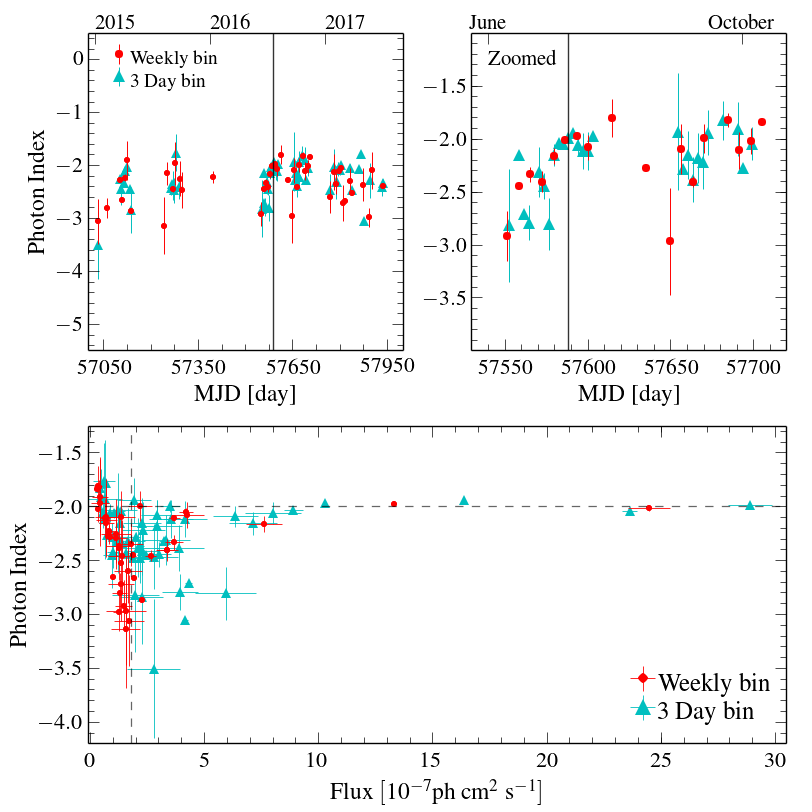}
 \caption{Evolution and distribution of the $\gamma$-ray photon index $\Gamma$ from the LAT data; red points mark the values for the weekly light curve, cyan points the values for the 3-day light curve. \emph{Top panels:} $\Gamma$ as function of time. Vertical solid lines indicate the $\gamma$-ray outburst on MJD 57588. The top left panel spans the entire time of our observations, the top right panels focuses on the period of high $\gamma$-ray activity in 2016; note the different axis scales. \emph{Bottom panel:} $\Gamma$ as funcion of $\gamma$-ray flux. The dashed horizontal and vertical lines represent a photon index of $\Gamma=-2$ and a flux of $1.8\times10^{-7}$ ph cm$^{-2}$ s$^{-1}$, respectively.}
 \label{fig:f3}
\end{figure}

\subsection{Multi-wavelength flux correlations}
\label{sec:resB} 

A physical connection between the emission at various wavelengths is apparent already from the morphology of the light curves (cf. Figure~\ref{fig:f1}). For a more quantitative analysis, we computed Pearson ($r_p$) and Spearman rank ($r_s$) correlation coefficients to probe the degrees of correlation between the $\gamma$-ray light curve and the emission at lower energy bands. We included all data from the period of enhanced $\gamma$-ray activity in mid-to-end 2016 that are simultaneous with the $\gamma$-ray data within the bin size of three days. For the optical data, we used flux estimates in linear units, in mJy, provided by the ASAS-SN online database along with the V band magnitudes. The OVRO 15 GHz data represent the radio band in the correlation analyis; the other radio light curves did not provide enough simultaneous data points. We assume false alarm probabilities $p\leq0.05$ to indicate statistically significant correlations (e.g., \citealt{Leung2014}). Figure~\ref{fig:f2} shows the results of the correlation analysis. All correlation coefficients, with values $r_p\geq0.69$ and $r_s\geq0.75$, point toward strong positive correlations between emission at $\gamma$-rays and at lower frequencies. False alarm probabilities are lower than 0.05 with the (marginal) exception of $p_{s}=0.052$ for the X-ray--$\gamma$-ray pair.

\subsection{LAT $\gamma$-ray photon indices}
\label{sec:resC} 

\begin{figure*}
 \includegraphics[width=\textwidth]{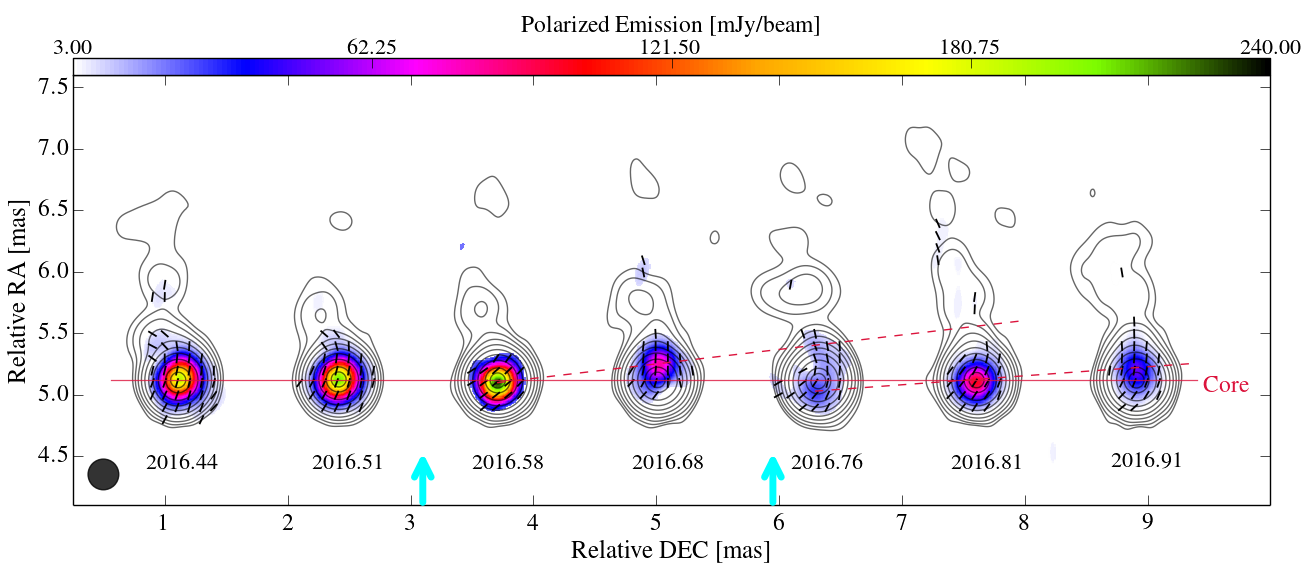}
 \caption{Linearly polarized flux density (\textit{color scale}), EVPA (\textit{black line segments}), and total intensity structure (\textit{background contours}) of 1749+096 at 43\,GHz as observed with the VLBA; from left to right: 2016 June 11, 2016 July 5, 2016 July 31, 2016 September 5, 2016 October 6, 2016 October 23, and 2016 November 28. The cyan arrows at the declination axis indicate the times of the two major $\gamma$-ray events: the 2016 July 19 outburst and the 2016 October 2 maximum. All maps are restored with a circular beam with a size of 0.25\,$\times$\,0.25~mas (displayed at the bottom left). Contour levels increase by factors of two from 0.25\% to 64\%, plus one additional countour corresponding to 85\% of the maximum total intensity. The two tilted red dashed lines indicate the apparent motion of the peak of the \emph{polarized} emission. The horizontal red solid line indicates the location of the 7-mm core.}
 \label{fig:f4}
\end{figure*}

We quantify the $\gamma$-ray spectrum of 1749+096 using the photon index $\Gamma$, which is defined as $dN/dE \propto E^{+\Gamma}$ with $N$ being the number of photons and $E$ being the photon energy. Figure~\ref{fig:f3} shows the photon indices obtained from the 3-day binned and weekly binned LAT light curves at 0.1--300\,GeV as function of time and as fuction of $\gamma$-ray flux, respectively. The photon index varies from $-3.5$ to $-1.7$ during the time of our observations; the two-year average value is $\Gamma=-2.3$. The photon index time series indicates a spectral hardening (i.e., an increase of $\Gamma$) around the time of the 2016 July $\gamma$-ray outburst. More specifically, the photon indices increase from about $-3$ to about $-2$ during the $\sim$40 days before the $\gamma$-ray outburst.
The photon index appears to decrease (from $-1.7$ to $-3.5$) with increasing flux until the $\gamma$-ray flux reaches about $\sim1.8\times10^{-7}$ ph cm$^{-2}$ s$^{-1}$. With further increasing flux the photon index increases again and approaches a plateau at $\Gamma\approx2$ for fluxes $\gtrsim6\times10^{-7}$ ph cm$^{-2}$ s$^{-1}$. 

\begin{table}
 \caption{Properties of the polarized VLBA component}
 \label{tab:tb1}
 \centering 
 \begin{tabular}{c @{\hspace{0.5cm}} c @{\hspace{0.2cm}} c @{\hspace{0.2cm}} c @{\hspace{0.2cm}} c}
  \hline		
Date & \textit{m}$_\mathrm{total}$ & PI$_\mathrm{peak}$ & EVPA$_\mathrm{peak}$ & rms$^a$\\
  &  (\%)  &  (mJy/beam)  &  ($^{\circ}$)  &  (mJy/beam)  \\
  \hline
2016 June 11  &  5.9  &  189  &  $-$14  &  0.69   \\
2016 July 05  &  4.5  &  208  &  $-$19  &  0.54   \\
2016 July 31  &  4.9  &  232  &  $-$51  &  2.73   \\
2016 Sept 05  &  2.4  &  84  &  $-$3  &  0.97   \\
2016 Oct 06  &  1.3  &  30  &  $-$39  &  1.01   \\
2016 Oct 23  &  5.0  &  102  &  $-$32  &  0.59   \\
2016 Nov 28  &  3.6  &  51  &  $-$10  &  0.42   \\
  \hline
  \multicolumn{5}{l}{$^a$ rms noise of residual polarization map.}\\
 \end{tabular}
\end{table}


\begin{figure}
 \includegraphics[width=\columnwidth]{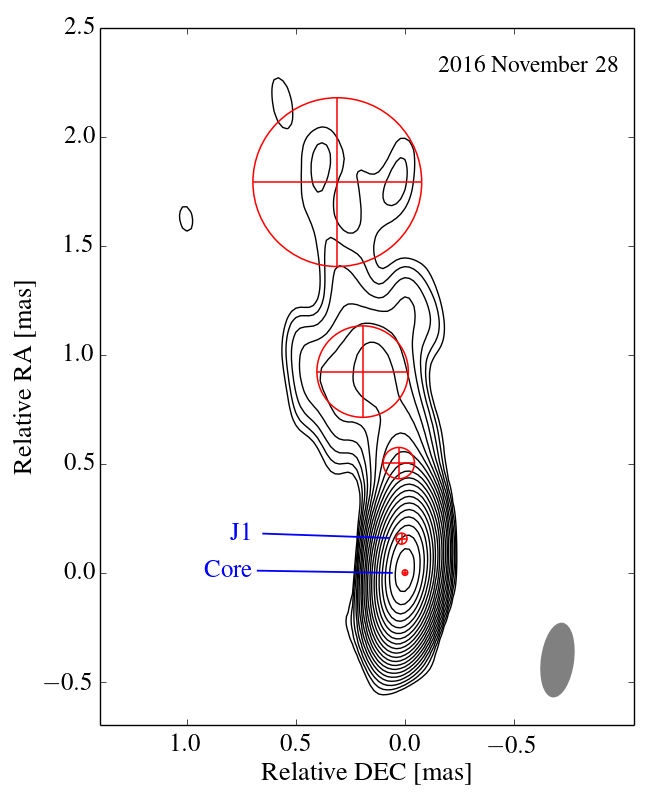}
 \caption{VLBA map of 1749+096 observed on 2016 November 28, from model component fitting. Individual circular Gaussian components are marked with red $\oplus$. The beam size (illustrated at the bottom right) is $0.34\times0.15$ mas at $-$7.6$^{\circ}$. Contour levels increase by factors of $\sqrt{2}$ from 0.11\% to 79.32\% of the total intensity peak. Blue solid lines point to the VLBA 43-GHz core and the jet component J1, respectively.}
 \label{fig:f5}
\end{figure}

\begin{figure}
 \includegraphics[width=\columnwidth]{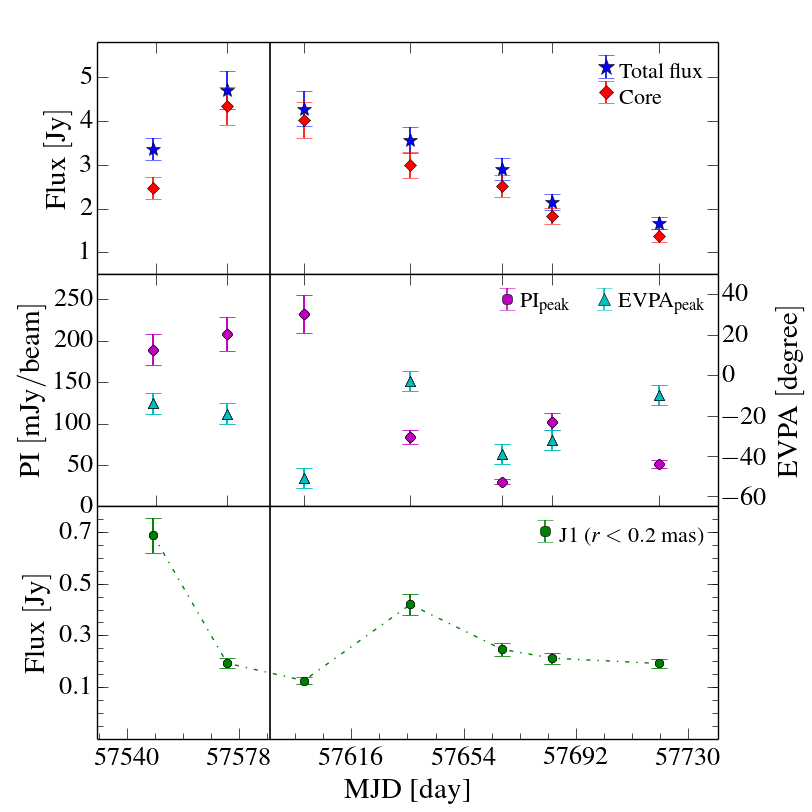}
 \caption{Evolution of flux density and the polarized component of 1749+096 obtained from the BU data; 2016 June 11 -- 2016 November 28. \emph{Top panel:} Core flux and total source flux. \emph{Middle panel:} PI and EVPA listed in Table~\ref{tab:tb1} with error of 10\% and 5$^{\circ}$, respectively. \emph{Bottom panel:} Evolution of the jet component J1, located at $\sim0.2$ mas from the core. The vertical solid line marks the time of the 2016 July $\gamma$-ray outburst.}
 \label{fig:f6}
\end{figure}

\subsection{Linear polarization at 43 GHz}
\label{sec:resD} 

Figure~\ref{fig:f4} shows the 7-mm (43\,GHz) linear polarization in the innermost few parsecs of the radio jet of 1749+096. Polarized intensity scale and EVPA markers are plotted over the total intensity contour. We applied an $8\sigma$ cutoff to the polarized intensity. We calculate fractional linear polarizations (defined as \textit{m} = $\sqrt{Q^{2} + U^{2}} / I$, where $I$, $Q$, and $U$ are Stokes parameters) by using CLEANed fluxes. We estimate a typical thermal noise of $\sim$1\,mJy/beam in the polarized emission. \citet{Jorstad2005} suggests typical systematic errors of the BU data of $\sim$1\% in fractional polarization and $\sim$5$^{\circ}$ in polarization angle for bright components, up to around 10$^{\circ}$ in the worst case (see also \citealt{Roberts1994} for discussion of uncertainties in the measured polarization). On average, we found a fractional linear polarization of $\sim$3.9\% throughout the BU dataset. Our further analysis focused on monitoring the properties of the polarized source component, corresponding to the peak of the polarized emission, which usually overlaps with the VLBA core. We summarize the polarization properties of 1749+096 in Table~\ref{tab:tb1}. Through 2016 June and 2016 July, the polarized flux increases notably. After the $\gamma$-ray outburst on 2016 July 19, the polarized intensity reaches its highest value, $\sim$230 mJy, and displays a rotation of the EVPA by about 32$^{\circ}$ with respect to the previous epoch. In 2016 September, the polarized component had moved down the jet, with the EVPA being aligned closely to the jet axis which is located at a position angle of $\sim$7.5$^{\circ}$ (the average position angle of all the jet components in Table~\ref{tab:tb2}). A new polarized component emerged \emph{upstream} of the VLBA core just four days after the 2016 October 2 $\gamma$-ray flux maximum.

\subsection{Flux evolution near the core}
\label{sec:resE} 

In addition to the polarized flux, the BU data provide important information on the (total intensity) structure of the innermost region of 1749+096. In order to trace the flux evolution of the various source components, we fit circular Gaussian profiles to them. The fitted parameters of all components are displayed in Table~\ref{tab:tb2}. Despite the clear evolution of the location of the polarized flux near the core (Section \ref{sec:resD}), we do not find indication for the ejection of a new jet component. As we might have missed a new component due to insufficient angular resolution or a smaller Doppler factor (i.e., a change of orientation in the curved jet), we took a closer look at the behaviour of the jet component J1. This component is located within 0.2 mas from the core in all 43-GHz VLBA maps (see Figure~\ref{fig:f5} and Table~\ref{tab:tb2}). Figure~\ref{fig:f6} shows the flux evolution of the core, J1, the polarized component, and total integrated source flux. The total flux is dominated by the core which contributes about 86\% of the total observed flux on average. The core flux peaks at the time of the 2016 July $\gamma$-ray outburst, thus suggesting the core as the origin of the radio flux counterpart to the $\gamma$-ray flare. In 2016 September, the flux of the component J1 increases by a factor of about 3 compared to 2016 July, a bit more than one month after the 2016 July $\gamma$-ray outburst.

\begin{table*}
 \caption{Parameters of the model fitted jet components in the 43\,GHz total intensity images.}
  \label{tab:tb2}
 \begin{tabular*}{\textwidth}{cccccccccccc}
  \hline
   \hline
Date & MJD & ID$^a$ & Flux & Distance$^b$ & Angle$^c$ & Diameter & B$_\mathrm{maj}^d$ & B$_\mathrm{min}^d$ & B$_\mathrm{PA}^d$ & rms$^e$ \\
 & & & (Jy) & (mas) & ($^{\circ}$) & (mas) & (mas) & (mas) & ($^{\circ}$) & (mJy/beam) \\
 \hline
  \hline
2016 June 11 & 57550 & C & 2.47 & 0.00 & $-$ & 0.02 & 0.35 & 0.15 & $\mathrm{-}$6.33 & 1.02 \\
 & & J1 & 0.69 & 0.09 & $\mathrm{-}$12.69 & 0.04 &  &  &  & \\
  & & J2 & 0.09 & 0.39 & 22.58 & 0.16 &  &  &  & \\
   & & J3 & 0.03 & 0.81 & 7.94 & 0.14 &  &  &  & \\
    & & J4 & 0.08 & 1.43 & 9.16 & 0.67 &  &  &  & \\
 \hline
2016 July 05 & 57574 & C & 4.35 & 0.00 & $-$ & 0.03 & 0.40 & 0.16 & $\mathrm{-}$12.60 & 1.09 \\
 & & J1 & 0.19 & 0.14 & $\mathrm{-}$0.11 & 0.05 &  &  &  & \\
  & & J2 & 0.09 & 0.55 & 15.48 & 0.21 &  &  &  & \\
   & & J3 & 0.07 & 1.39 & 7.81 & 0.64 &  &  &  & \\
 \hline
2016 July 31 & 57600 & C & 4.02 & 0.00 & $-$ & 0.03 & 0.35 & 0.17 & $\mathrm{-}$7.63 & 1.41 \\
 & & J1 & 0.12 & 0.17 & 0.30 & 0.04 &  &  &  & \\
  & & J2 & 0.06 & 0.53 & 15.15 & 0.17 &  &  &  & \\
   & & J3 & 0.03 & 0.79 & 13.23 & 0.10 &  &  &  & \\
    & & J4 & 0.05 & 1.46 & 2.73 & 0.43 &  &  &  & \\
 \hline
2016 Sept 05 & 57636 & C & 2.99 & 0.00 & $-$ & 0.02 & 0.49 & 0.17 & $\mathrm{-}$20.40 & 1.44 \\
 & & J1 & 0.42 & 0.10 & 11.65 & 0.03 &  &  &  & \\
  & & J2 & 0.05 & 0.43 & $\mathrm{-}$0.53 & 0.08 &  &  &  & \\
   & & J3 & 0.07 & 0.69 & 15.21 & 0.22 &  &  &  & \\
    & & J4 & 0.04 & 1.54 & 2.12 & 0.41 &  &  &  & \\
 \hline
2016 Oct 06 & 57667 & C & 2.51 & 0.00 & $-$ & 0.05 & 0.63 & 0.20 & $\mathrm{-}$28.00 & 1.48 \\
 & & J1 & 0.25 & 0.15 & $\mathrm{-}$2.75 & 0.08 &  &  &  & \\
  & & J2 & 0.10 & 0.73 & 10.75 & 0.32 &  &  &  & \\
   & & J3 & 0.04 & 1.78 & 9.65 & 0.76 &  &  &  & \\
 \hline
2016 Oct 23 & 57684 & C & 1.83 & 0.00 & $-$ & 0.02 & 0.35 & 0.15 & $\mathrm{-}$3.96 & 0.99 \\
 & & J1 & 0.21 & 0.13 & 5.94 & 0.04 &  &  &  & \\
  & & J2 & 0.03 & 0.45 & 10.39 & 0.14 &  &  &  & \\
   & & J3 & 0.04 & 0.87 & 11.40 & 0.21 &  &  &  & \\
    & & J4 & 0.04 & 1.69 & 8.76 & 0.55 &  &  &  & \\
 \hline
2016 Nov 28 & 57720 & C & 1.38 & 0.00 & $-$ & 0.03 & 0.34 & 0.15 & $\mathrm{-}$7.63 & 0.53 \\
 & & J1 & 0.19 & 0.16 & 5.57 & 0.05 &  &  &  & \\
  & & J2 & 0.02 & 0.50 & 3.15 & 0.14 &  &  &  & \\
   & & J3 & 0.05 & 0.94 & 11.85 & 0.42 &  &  &  & \\
    & & J4 & 0.03 & 1.82 & 9.81 & 0.77 &  &  &  & \\
 \hline
  \hline
\multicolumn{11}{l}{$^a$ Higher ID numbers Jx correspond to larger downstream distances from the core.}\\
\multicolumn{11}{l}{$^b$ Distance from the core.}\\
\multicolumn{11}{l}{$^c$ Position angle relative to core component.}\\
\multicolumn{11}{l}{$^d$ Parameters of the elliptical beam: major axis, minor axis, position angle}\\
\multicolumn{11}{l}{$^e$ rms noise of residual map.}\\
 \end{tabular*}
\end{table*}

\section{Discussion}

\subsection{$\gamma$-ray activity}
\label{sec:disA} 

The 2016 July $\gamma$-ray outburst is an exceptional event with no precedent since the beginning of $\gamma$-ray observations in 2009. During phases of low as well as very high $\gamma$-ray fluxes, we observe photon indices close to the value $\Gamma\approx-2.2$ expected for LSP blazars, suggesting a spectral break located at around 100 MeV or less \citep{Lico2017}. Else than intermediate-synchrotron peaked (ISP) and high-synchrotron peaked (HSP) blazars, LSP blazars are known to experience severe cooling in the energy range 0.1--300\,GeV (i.e, our LAT band), thus producing the IC component of the SED \citep{Lico2017}. During the 2016 July $\gamma$-ray outburst we observe an increase in photon index (see Figure~\ref{fig:f3}), meaning a spectral hardening with increasing flux (e.g., \citealt{Nandikotkur2007}). Such behavior is rare in BL Lac objects \citep{Lico2014}. \citet{Kushwaha2014} suggested shock acceleration as explanation for the hardening of $\gamma$-ray spectra. In our case, the temporal agreement of the apex of the spectral hardening with the $\gamma$-ray flux peak points toward shock acceleration inducing a surge of $\gamma$-ray photons at higher energies efficiently \citep{Kusunose2000}.

\citet{Abdo2010} found a transition between a harder-when-brighter trend and a softer-when-brighter trend in PKS 1510$-$089 for energies above 0.2 GeV. In their observation, photon indices softened with fluxes increasing up to $\sim2.4\times10^{-7}$ ph cm$^{-2}$ s$^{-1}$, then hardened again with fluxes increasing further. This matches our observation of decreasing photon index with the flux increasing up to about $1.8\times10^{-7}$\,ph\,cm$^{-2}$\,s$^{-1}$ (weekly light curve, see Figure~\ref{fig:f3}). Assuming a threshold value of $1.7\times10^{-7}$ ph cm$^{-2}$ s$^{-1}$, we find a strong negative correlation ($r_p=-0.86$) between $\gamma$-ray flux and $\Gamma$, corresponding to a softer-when-brighter scaling. The physical mechanisms behind this softer-when-brighter trend as well as the inversion of this trend at a certain threshold flux are unclear \citep{Abdo2010}. Candidate mechanisms are a change in emission mechanism (e.g., \citealt{Asano2014}), the cooling time scale (e.g., \citealt{Dotson2012}), or the magnetic field strength (e.g., \citealt{Kusunose2000}).

\subsection{Multi-wavelength correlations}
\label{sec:disB} 

The $\gamma$-ray outburst in 2016 July was accompanied by simultaneous flux enhancements from radio to X-rays, indicating a physical connection across all wavelengths \citep{Jorstad2001a, Tavares2011, Lico2017}. The fact the source flux from radio to $\gamma$ peaks simultaneously within few days strongly suggests that the emission at all wavelengths (largely) originates from the same location within the source \citep{Tavares2011, Wehrle2012, Jorstad2013, Casadio2015}. A possible exception is the 15-GHz radio peak which we cannot locate exactly and which might be delayed by up to a few more days relative to the events at higher energy bands. If it were indeed delayed, we would have to assume a displacement of the radio emitting region relative to the regions emitting higher energy radiation. This is supposed to occur when the emission at higher energies is produced in a region that is optically thick at 15\,GHz; radio light is emitted only once the disturbance in the jet has entered a region transparent at 15\,GHz (\citealt{Wehrle2012}; see also \citealt{Agudo2011}, for discussion of a physically extended disturbance).

Figure~\ref{fig:f6} shows that the BU VLBA core is the origin of the 7-mm outburst, thus implying a connection between the mm-wave core and the simultaneous events at higher energy bands (e.g., \citealt{Agudo2011, Wehrle2012, Jorstad2013}). The peak of the $\gamma$-ray outburst seems to coincide with the peak of the 7-mm emission. This is unexpected as the conventional picture of the radio--$\gamma$-ray connection expects a $\gamma$-ray outburst at the onset (during the rise) of a radio flare (\citealt{Marscher2016}; but see also \citealt{Valtaoja1995, Moerbeck2014}, for various timings of $\gamma$-ray events relative to radio flares). However, we note that a considerable number of studies observed $\gamma$-ray outbursts being (quasi-)simultaneous with radio flares (e.g., \citealt{Tavares2011, Wehrle2012, Lico2014, Casadio2015}).

\subsection{Origin of the $\gamma$-ray outburst}
\label{sec:disC} 

The relative timing of the $\gamma$-ray outburst and the 7-mm outburst of the VLBA core suggests the mm-wave core to be the production site of the $\gamma$-radiation \citep{Wehrle2012, Jorstad2013, Casadio2015}. The behaviour of the polarized VLBA component after the 2016 July $\gamma$-ray outburst supports this idea (see Figure~\ref{fig:f4}). The linear polarization image of 2016 September 5 shows clearly that the region of polarized emission, which was located at the VLBI core before, moved down the jet. This can be interpreted as the signature of a shock emerging within, and moving away from, the core (e.g., \citealt{Ros2000, Marscher2008, Pushkarev2008, Marscher2010, Agudo2011, Wehrle2012, Marscher2012, Jorstad2013}). This picture connects the $\gamma$-ray outburst with the passage of a propagating disturbance, like a new jet component, through the core. The flux evolution of component J1, that was detected within 0.2\,mas from the core consistently over six months, supports this idea (see Figure~\ref{fig:f6}). The enhancement of the flux from J1 as observed on 2016 September 5 is consistent with the disturbance propagating through a region downstream of the core \citep{Casadio2015, Hodgson2017}. However, we do not find a newly emerging feature in the (total intensity) VLBA maps that could be associated with displacement of the polarized component. Interestingly, \citet{Lico2014} and \citet{Ros2000} encountered similar situations in Mrk\,421 and 3C\,345. The absence of a directly observed new jet component may be attributed to a complex structure of the jet around the core or spatial blending of multiple emission regions in the jet \citep{Ros2000, Jorstad2013, Hodgson2017}.

As noted in Section~\ref{sec:disA}, the evolution of the $\gamma$-ray spectrum of 1749+096 around the time of the 2016 July $\gamma$-ray flare is consistent with the acceleration of a relativistic shock. Further evidence in favor of this interpretation is provided by the evolution of the BU VLBA core flux both in total intensity and linear polarization during 2016 June and 2016 July. The polarized flux reached its maximum on 2016 July 31, when the BU core flux was in decline already (see Figure~\ref{fig:f6}). This is the signature expected from a disturbance propagating through the core (e.g., \citealt{Lico2014}) but may also be connected to the evolution of a relativistic shock (\citealt{Ros2000}; see also \citealt{Tavares2012} for discussions of strong core polarization). The maximum of the polarized intensity, about 230\,mJy/beam, coincides with an EVPA swing by $\sim$32$^{\circ}$ in the core region on 2016 July 31. \citet{Hughes2011} suggested that initially random and turbulent magnetic fields in a blazar jet can be compressed by a propagating oblique shock, thus leading to both an enhancement of polarized intensity and a swing of the EVPA \citep{Laing1980, Hughes1985, Hughes1991}. Accordingly, we suggest that the passage of a propagating disturbance through the mm-wave core is responsible for the $\gamma$-ray outburst. \citep{Marscher2008, Agudo2011, Wehrle2012, Jorstad2013}. Further clues for constraining the production site of the optical-to-$\gamma$-ray outbursts could be provided by the opacity of the core at 7\,mm (see Section~\ref{sec:disD}) during the event. It seems that the $\gamma$-ray outburst is contemporaneous with its 7-mm counterpart, whereas the cm-wave counterpart in the OVRO light curve is slightly delayed relative to the peak of the $\gamma$-ray outburst. This leads us to consider a region downstream of the mm-wave core to be the origin of the $\gamma$-ray outburst. This is in agreement with the disturbance being spatially extended (e.g., \citealt{Agudo2011}). The duration of the $\gamma$-ray outburst, roughly 50 days, can be considered as the time needed for the disturbance to pass through the mm-wave core \citep{Jorstad2013}. Then, the strongest $\gamma$-ray emission might be produced by the back region of the propagating disturbance $\sim$10 days before the disturbance fully escapes the mm-wave core.

\subsection{The enhanced $\gamma$-ray emission in 2016 October}
\label{sec:disD} 
\begin{figure}
 \includegraphics[width=\columnwidth]{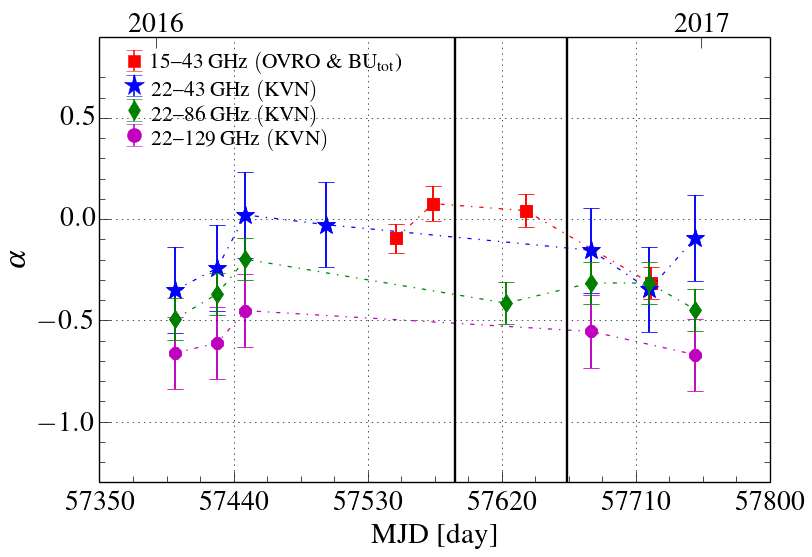}
 \caption{Pairwise spectral indices ($S_\nu \propto \nu^{\alpha}$) of 1749+096 at radio wavelengths observed in 2016. Different colors and symbols indicate different frequency pairs. The epochs of the $\gamma$-ray events are indicated by two black vertical solid lines.}
 \label{fig:f7}
\end{figure}
We find a notable $\gamma$-rays flux enhancement around 2016 October 2. Contrary to the prior (2016 July) $\gamma$-ray outburst, this local peak was, if at all, accompanied only by an optical counterpart without corresponding flux enhancements at radio and X-rays. The 43-GHz linear polarization maps obtained after the time of the $\gamma$-ray enhancement show a polarized component propagating down the jet from the BU core. This suggests the presence of a propagating disturbance similar to the situation discussed in Section~\ref{sec:disB}. Notably, we observe an upstream displacement of the polarized component relative to the peak of the total intensity on 2016 October 6. However, the data from this epoch need to be interpreted with care owing to reduced sensitivity and resolution caused by antennas missing from the array (see Section~\ref{sec:obsA}). We calculated spectral indices of 1749+096 from the radio data as follows. For the pair of 15--43\,GHz, we employed OVRO and the BU total fluxes observed within 1 day, assuming that the VLBI-scale structure of the source dominates the OVRO fluxes (the ratio of the OVRO and BU fluxes is 1.1 on average). Although the data points are sparse, it seems that the source was opaque ($\alpha$~$\sim$~0) at 43\,GHz during the $\gamma$-ray flaring period. This is consistent with what \citet{osullivan2009} reported (i.e., a spectral index of the core region of 1749+096 $\sim-0.1$ between 12.9 and 43\,GHz). We consider that the second $\gamma$-ray event might have been caused by a propagating disturbance at (nearly) the same location in the jet as the first $\gamma$-ray event, but with smaller Doppler factor and energization. This explains the relatively lower $\gamma$-ray flux density and the absence of a radio counterpart compared to the major $\gamma$-ray outburst.

\section{Conclusions}
In this study, we explored the nature of two $\gamma$-ray events, the 2016 July outburst and the 2016 October flux enhancement, in the blazar 1749+096. From the combined evidence provided by multi-waveband flux observations plus 43-GHz VLBA maps, we conclude that both $\gamma$-ray events are connected to the propagation of a disturbance in the jet \citep{Jorstad2001b, Marscher2008, Jorstad2016}. Regarding the origin of the two $\gamma$-ray events, we suggest the `parsec-scale scenario' (e.g., \citealt{Agudo2011, Wehrle2012, Jorstad2013}) where a relativistic shock moving down the jet causes an enhancement of $\gamma$-ray flux in the radio core by providing highly accelerated electrons. As the disturbance passes through the mm-wave core, the VLBA core flares simultaneously with the fluxes at the higher energy bands. Eventually, a moving feature can be seen in the linear polarization images. Given that the cm-wave flux peak is slightly delayed (around 5 days) relative to the $\gamma$-ray outburst, we tentatively conclude that a region downstream of the mm-wave core is the origin of the $\gamma$-ray outburst. The $\gamma$-ray outburst matches the growth of a strong shock. We find a hardening of the $\gamma$-ray spectrum with increasing flux during the rising stage of the $\gamma$-ray outburst. The subsequent presence of abundant polarized emission in the core region further supports the presence of growing shock \citep{Tavares2012}.

In the case of the $\gamma$-ray enhancement on 2016 October 2, we find an upstream displacement of the polarized peak relative to the total intensity in the linear polarization image on 2016 October 6. Given the opacity of 1749+096 at 43\,GHz, however, the polarized component cannot be detected upstream of the BU core due to synchrotron self-absorption. For this $\gamma$-ray event, we expect that the event was less energized with relatively smaller Doppler factor, thus resulting in some differences in the observed evolution between the two $\gamma$-ray events.

The origin of the bulk of the seed photons remains unclear for both $\gamma$-ray events. In general, both the internal IC process with seed photons from the jet itself (\citealt{Marscher2010}), and external Compton (EC) scattering with seed photons from the dusty torus at parsec-scales or the BLR at subparsec-scales (see also \citealt{Tavares2011}, for discussion of outflowing BLR at parsec-scales) can be considered. Given the observed hardening of the $\gamma$-ray spectrum however, the EC process with infrared seed photons from the dusty torus might be the dominant emission mechanism for the 2016 July $\gamma$-ray outburst \citep{Agudo2011}. A better understanding of $\gamma$-ray flares in blazar jets not only requires monitoring the properties of the linear polarization (which reflects the underlying magnetic field configuration; e.g., \citealt{Homan2002, Marscher2012}), but also changes in jet component Doppler factors caused by viewing angle variations that could substantially affect the observed $\gamma$-rays \citep{Jorstad2001b, Casadio2015, Raiteri2017}.

\section*{Acknowledgements}
We appreciate the referee for constructive and fruitful comments which improved the manuscript. We thank R. Prince (RRI), J. Perkins (NASA/GSFC), and R. Corbet (UMBC/NASA) for their valuable advice on the analysis of the LAT data. We thank E. Ros (MPIfR) for useful comments. We are grateful to the staff of the KVN who helped to operate the array and to correlate the data. The KVN and a high-performance computing cluster are facilities operated by the KASI (Korea Astronomy and Space Science Institute). The KVN observations and correlations are supported through the high-speed network connections among the KVN sites provided by the KREONET (Korea Research Environment Open NETwork), which is managed and operated by the KISTI (Korea Institute of Science and Technology Information). This study makes use of 43 GHz VLBA data from the VLBA-BU Blazar Monitoring Program (VLBA-BU-BLAZAR; \url{http://www.bu.edu/blazars/VLBAproject.html}), funded by NASA through the Fermi Guest Investigator Program. The VLBA is an instrument of the Long Baseline Observatory. The Long Baseline Observatory is a facility of the National Science Foundation operated by Associated Universities, Inc. This research has made use of data from the OVRO 40-m monitoring program \citep{Richards2011} which is supported in part by NASA grants NNX08AW31G, NNX11A043G, and NNX14AQ89G and NSF grants AST-0808050 and AST-1109911. This work made use of data supplied by the UK Swift Science Data Centre at the University of Leicester. We acknowledge financial support by the National Research Foundation of Korea (NRF) via research grants 2015-R1D1A1A01056807 (D. Kim, S. Trippe, JC. Algaba), 2016-R1C1B2006697 (S. Lee, S. Kang), 2015-H1D3A1066561 (G. Zhao), and 2014-H1A2A1018695 (J. Park). J. W. Lee is grateful for the support of the National Research Council of Science and Technology, Korea (Project Number EU-16-001).





\bsp	
\label{lastpage}
\end{document}